\documentclass[aps,prb,preprint,superscriptaddress,showkeys]{revtex4-1}
\usepackage{amsmath,amsfonts,amssymb,bm}
\newcommand{\abs}[1]{\left\vert#1\right\vert}
\numberwithin{equation}{section}
\begin{document}
\title{Path integrals of a particle in a finite interval and on the half-line}
\author{Seiji Sakoda}
\email{sakoda@nda.ac.jp}
\affiliation{Department of Applied Physics, National Defense Academy,
Hashirimizu\\
Yokosuka city, Kanagawa 239-8686, Japan}

\date{\today}

\begin{abstract}
We make use of point transformations to introduce new canonical variables for systems defined on a finite interval and on the half-line so that new position variables should take all real values from $-\infty$ to $\infty$. The completeness of eigenvectors of new momentum operators enables us to formulate time sliced path integrals for such systems. Short time kernels thus obtained require extension of the range of variables to the covering space in order to take all reflected paths into account. Upon this extension we determine phase factors attached to the amplitude for paths reflected at boundaries by taking singularities of the potential into account. It will be shown that the phase factor depends on parameters that characterize the potential and further that the wellknow minus sign in the amplitude for odd times reflection of the free particle in a box should be understood as the special case for the corresponding value of the parameter of the potential.
\end{abstract}


\maketitle

\section{Introduction}
Since the early days of quantum mechanics there exists only a limited set of exactly solvable models such as the harmonic oscillator and hydrogen-like potentials, the square well potential, the Morse potential and a fews. The factorization method, introduced by Schr\"{o}dinger\cite{Schrodinger46A9,Schrodinger46A183,Schrodinger47A} and Dirac\cite{Dirac}, opened a systematic treatment for dealing with systems of exact solvability. The classification of types in factorization of the second order differential equations was given by Infeld and Hull\cite{Infeld_Hull}. Today these solvable systems are well studied and understood from the viewpoint of shape invariance\cite{Gendenshtein} or supersymmetric quantum mechanics\cite{Witten1981,DKS,CKS,CKSbook}. When utilized in conjunction with Crum's theorem\cite{Crum}, shape invariance or the factorization of a Hamiltonian ensures the exact solvability of the model in the Schr\"{o}dinger picture.

In the framework of the Heisenberg picture, the solvability of a one-dimensional quantum mechanical system is related to the existence of a sinusoidal coordinate. If we can find a sinusoidal coordinate, by introducing creation and annihilation operators as its negative and positive frequency part respectively, we can define the ground state to be destroyed by the annihilation operator; then excited states are generated by repeated multiplication of the creation operator. The existence of a sinusoidal coordinate is connected to the closure relation\cite{Odake_Sasaki} which explicitly means that the second derivative, when differentiated by time, of a sinusoidal coordinate must be a linear combination of its first derivative as well as itself. Including the harmonic oscillator, there exist several models those can be solved by means of the algebraic method basing upon the creation and annihilation operators mentioned above but not all the systems with shape invariance can be solved by this method. 

Considering the solvability of one-dimensional quantum systems from the path integral point of view, it had been restricted for a long time to systems described by quadratic Lagrangians defined over the unrestricted one-dimensional space. The breakthrough was the solvability of the path integral for hydrogen atom by means of the Duru-Kleinert(DK) transformation\cite{DK79,DK}. Initiated by their brilliant work, there appeared many works as the application of their formalism\cite{Inomata82,HoInomata,Inomata,RingwoodDevreese,BlanchardSirugue,Steiner,PakSokmen,Duru86,ChetouaniChetouani,YoungDeWitt-Morette,Kleinert87,Bohm_Junker,CastrigianoStaerk,KleinertBook,Kleinert98}. Also some attempts were made to elaborate on the meaning of the method and the gauge invariance behind it\cite{Fujikawa,Sakoda,Sakoda2017}. In this way, the number of solvable path integrals has been considerably increased with the aid of the DK transformation. In addition to path integrals exactly solvable by means of DK transformation, there have been found path integrals defined on some homogeneous spaces to be evaluated exactly by semiclassical method, thus providing examples of the localization formula\cite{Rajeev1994,FKSF95_3232,FKSF95_4590,FKNS95,FKS96}. The key to understand the solvability of these systems with the localization formula was the existence of the quadratic Hamiltonian with first class constraints.
Another kind of exactly solved path integration will be those for the free particle in a box or on a circle\cite{Schulman68,Pauli1974,Laidlaw_DeWitt,DeWitt_Morette,Janke_Kleinert,Marinov80,Inomate_Singh}. The Feynman kernel for a particle on a circle is characterized by the emergence of one dimensional representation of the fundamental group basing upon the analysis from the viewpoint of homotopy while the phase factor $-1$ upon each reflection at boundaries being required for the Feynman kernel of the free particle in a box. If the phase factor is universal and determined purely by geometric argument, the same factor will also appear in the path integral for a particle in a box with a potential and the Feynman kernel may have the decomposition as a sum over paths with this geometric phase factor.
 
Common to exactly solvable path integrals mentioned above, excepting systems with quadratic Lagrangians on a free one-dimensional space and a particle on a circle, the origin of the difficulty in formulating path integrals would be that these systems are defined on a finite interval or on the half-line upon which we cannot define Hermitian momentum operators. Nevertheless, in many papers, path integrals for these systems are argued in terms of the Lagrangian path integral; and moreover most of them are expressed in the continuum representation. As is well known, Lagrangian path integrals for the harmonic oscillator and a free particle are obtained from the Hamiltonian one after performing the Gaussian integration with respect to the momentum. Furthermore the continuum representation of a path integral can be defined just as a limit from a time sliced one formulated with due care. Absence of the Hermitian momentum operator prevents us from carrying out the procedure of the transition from a Hamiltonian path integral to the corresponding Lagrangian form in most cases excepting the use of radial plane wave for the radial path integral\cite{Fujikawa2008}. Regarding the path integral on the half-line, there seems to be an attempt to formulate it from the viewpoint of random walk\cite{CMS1980} by modifying the action to incorporate the effect of the boundary condition into as well as the one to define the measure of path integral in a suitable way according to the time paths spend near the boundary\cite{Farhi_Gutmann}. These approaches cannot be, however, generic prescriptions for constructing path integrals of solvable systems. Therefore, including these rather special methods for path integrals on the half-line, Lagrangian path integrals in the continuum representation for salvable models must be validated by testing the consistency with the Schr\"{o}dinger equation.  In view of these facts, there seems to remain a room to put time sliced path integrals to be examined for solvable systems constructed from the Hamiltonian formulation by keeping rigid connection with the Schr\"{o}dinger theory. We shall try in this paper to develop a method of conversion from Hamiltonian path integrals into Lagrangian ones for systems on a finite interval and on the half-line. To this aim, we introduce a point transformation\cite{Gervai_Jevicki,Omote1977,FukutakaKashiwa,Ohnuki_Watanabe2003} from the original position variable $x$ to $Q=Q(x)$, which is suitably defined to take all real values from $-\infty$ to $\infty$, for each case. In addition to this new position variable, we define its canonical conjugate $P$ as a new momentum operator. The completeness of eigenvectors of new momentum operator defined this way will be utilized to formulate Hamiltonian path integrals and convert them into the Lagrangian ones.

This paper is organized as follows: in section 2 we describe the point transformation and see the completeness of the newly defined momentum eigenvectors. Section 3 gives explanation on the formulation of path integrals in terms of the eigenvectors of the new momentum operator. A brief sketch of a technique for evaluating the complicated kinetic term in the time sliced path integral thus formulated will also be given there. We performe explicit calculations of time sliced path integrals of Feynman kernels for symmetric and generalized P\"{o}schl-Teller potential as well as the radial harmonic oscillator in section 4. The final section is devoted to the conclusion. The ambiguity in the form of the potential term in the path integral will be briefly discussed in the appendix.

\section{Point transformation to new canonical variables}
For a system defined on $a<x<b$, we introduce a monotonically increasing function $f(x)$ and define
\begin{equation}
\label{eq:defineQ}
	Q(x)\equiv\int_{x_{0}}^{x}\!\!
	\frac{\ dx\ }{\ f{'}(x)\ },\quad
	f{'}(x)\equiv\frac{\ df(x)\ }{\ dx\ },
\end{equation}
where $x_{0}$ should be chosen in a suitable way.
We further introduce a differential operator\cite{Ohnuki_Watanabe2003} $P$ by
\begin{equation}
\label{eq:defineP}
	P\psi(x)\equiv-\frac{\ i\hbar\ }{\ 2\ }
	\left\{f{'}(x)\frac{\ d\ \ }{\ dx\ }+
	\frac{\ d\ \ }{\ dx\ }f{'}(x)\right\}\psi(x)
\end{equation}
in addition to $Q(x)$ above.
An eigenvalue equation $P\psi_{P{'}}(x)=P{'}\psi_{P{'}}(x)$ can be solved by
\begin{equation}
	\psi_{P{'}}(x)=\frac{1}{\ \sqrt{2\pi\hbar f{'}(x)\,}\ }
	e^{iP{'}Q(x)/\hbar}.
\end{equation}
If the range of $Q(x)$ covers all real values from $-\infty$ to $\infty$,
we find that $\psi_{P}(x)$ satisfies the normalization
\begin{equation}
\begin{aligned}
	\int_{a}^{b}\!\!\psi_{P_{1}}^{*}(x)\psi_{P_{2}}(x)\,dx=&
	\int_{a}^{b}\!\!\frac{dx}{\ 2\pi\hbar f{'}(x)\ }
	e^{-i(P_{1}-P_{2})Q(x)/\hbar}\\
	=&
	\int_{-\infty}^{\infty}\!\!\frac{dQ}{\ 2\pi\hbar\ }
	e^{-i(P_{1}-P_{2})Q/\hbar}\\
	=&
	\delta(P_{1}-P_{2})
\end{aligned}	
\end{equation}
for real values $P_{1}$ and $P_{2}$. To make $Q(x)$ take all values from $-\infty$ to $\infty$, $f{'}(x)$ must vanish at both endpoints if both $a$ and $b$ are finite and, if $b=\infty$, $f{'}(a)=0$ is required for finite $a$.
The completeness of the eigenfunction of $P$ can be easily checked as follows:
\begin{equation}
	\int_{-\infty}^{\infty}\!\!\psi_{P}(x)\psi_{P}^{*}(x{'})\,dP=
	\frac{1}{\ \sqrt{f{'}(x)f{'}(x{'})\,}\ }
	\delta(Q(x)-Q(x{'}))=\delta(x-x{'}).
\end{equation}
We may therefore write the eigenfunction $\psi_{P}(x)$ as $\psi_{P}(x)=\langle x\vert P\rangle$ to write the completeness above as
\begin{equation}
	\int_{-\infty}^{\infty}\!\!\vert P\rangle\langle P\vert\,dP=\bm{1},
\end{equation}
where $\bm{1}$ is the unity.
If we introduce a bra vector $\langle Q\vert$ by
\begin{equation}
	\langle Q\vert\equiv\sqrt{f{'}(x)\,}\
	\langle x\vert\
	\text{ in addition to its conjugate }\
	\vert Q\rangle=(\langle Q\vert)^{\dagger},
\end{equation}
the definition of the operator $P$ is converted into
\begin{equation}
	\langle Q\vert P=-i\hbar\frac{\ d\ \ }{\ dQ\ }\langle Q\vert.
\end{equation}
The completeness of $\vert Q\rangle$ is equivalent to that of $\vert x\rangle$:
\begin{equation}
	\int_{-\infty}^{\infty}\!\!\vert Q\rangle\langle Q\vert\,dQ=
	\int_{a}^{b}\!\!\vert x\rangle\langle x\vert\,dx=\bm{1}.
\end{equation}
In this way we see that operators $Q=Q(x)$ and $P$ defined above form a pair of canonical variables which fulfills the canonical commutation relation
$[Q,\,P]=i\hbar$. Completeness of the eigenvectors of these new canonical variables will be useful in formulating Duru-Klinert path integrals for some systems if the new position variable $Q(x)$ ranges all real values. In the defintion of $Q(x)$, given by \eqref{eq:defineQ}, we have just required $f{'}(x)>0$ for $x$ in $a<x<b$. It will be now clear that we should further assume that $f{'}(x)$ must vanish at both endpoints for systems on a finite interval and that $f{'}(a)=0$ for the case $a$ is finite and $b=\infty$ to make $Q(x)$ take all values from $-\infty$ to $\infty$.

\section{Conversion of the Hamiltonian and the construction of the path integral}
On an interval $a<x<b$, we may consider a quantum system described by the Schr\"{o}dinger equation
\begin{equation}
	\left\{-\frac{\hbar^{2}}{\ 2m\ }\frac{\ d^{2}\ \ }{\ dx^{2}\ }
	+V(x)\right\}\psi(x)=E\psi(x)
\end{equation}
with boundary conditions $\psi(a)=\psi(b)=0$.
Since wave functions must vanish at the endpoints, the formally defined momentum operator $p$ given by
\begin{equation}
	\langle x\vert p=-i\hbar\frac{\ d\ \ }{\ dx\ }\langle x\vert
\end{equation}
cannot be self-adjoint when $a$ and/or $b$ being finite.
For such cases, we cannot utilize the completeness of the eigenvector of the momentum operator. To avoid this inconvenient situation, we rewrite the Hamiltonian
\begin{equation}
\label{eq:hamiltonian01}
	H=\frac{1}{\ 2m\ }p^{2}+V(x)
\end{equation}
in terms of $P$ introduced in the previous section.
By a simple and straightforward manipulation, we obtain
\begin{equation}
	H=\frac{1}{\ 2m\ }\frac{1}{\ f{'}(x)\ }P^{2}\frac{1}{\ f{'}(x)\ }+
	\frac{\hbar^{2}}{\ 8m\ }\left[
	\left\{\frac{\ f{''}(x)\ }{\ f{'}(x)\ }\right\}^{2}-
	\frac{\ 2f{'''}(x)\ }{\ f{'}(x)\ }\right]+V(x).
\end{equation}
It will be useful here to notice that a commutation relation of a function $G(x)$ and the operator $P$ is given by
\begin{equation}
	[G(x),\,P]=i\hbar f{'}(x)G{'}(x).
\end{equation}

We now make use of the completeness of the eigenvector of $P$ to find a short time kernel
\begin{equation}
	K(x,x{'};\epsilon)=
	\langle x\vert\left(\bm{1}-\frac{\ \epsilon\ }{\ \hbar\ }H\right)
	\vert x{'}\rangle
\end{equation}
with the imaginary time $\epsilon=\beta/N$($\beta>0$) for $N\to\infty$. 
For the term proportional to $P^{2}$, we evaluate it as
\begin{equation}
	\langle x\vert\frac{1}{\ f{'}(x)\ }P^{2}\frac{1}{\ f{'}(x)\ }
	\vert x{'}\rangle=
	\frac{1}{\ f{'}(x)f{'}(x{'})\ }
	\int_{-\infty}^{\infty}\!\!P^{2}\langle x\vert P\rangle
	\langle P\vert x{'}\rangle\,dP
\end{equation}
and we may assume some ordering prescription in the evaluation of potential terms to find
\begin{equation}
	\langle x\vert\left\{
	\frac{\hbar^{2}}{\ 8m\ }\left[
	\left\{\frac{\ f{''}(x)\ }{\ f{'}(x)\ }\right\}^{2}-
	\frac{\ 2f{'''}(x)\ }{\ f{'}(x)\ }\right]+V(x)\right\}
	\vert x{'}\rangle=V_{\mathrm{eff}}(x,x{'})
	\int_{-\infty}^{\infty}\!\!\langle x\vert P\rangle
	\langle P\vert x{'}\rangle\,dP.
\end{equation}
The precise form of $V_{\mathrm{eff}}(x,x{'})$, the dependences on $x$ and $x{'}$ in particular, will depend on systems we are dealing with. Keeping this in mind, we exponentiate the $c$-number Hamiltonian to obtain a Gaussian integral
\begin{equation}
\begin{aligned}
	K(x,x{'};\epsilon)
	=&
	\int_{-\infty}^{\infty}\!\!
	\frac{dP}{\ 2\pi\hbar\sqrt{f{'}(x)f{'}(x{'})\,}\ }\\
	&\times
	\exp\left[\frac{\ i\ }{\ \hbar\ }P\left\{Q(x)-Q(x{'})\right\}-
	\frac{\ \epsilon\ }{\ \hbar\ }\left\{
	\frac{P^{2}}{\ 2mf{'}(x)f{'}(x{'})\ }+
	V_{\mathrm{eff}}(x,x{'})
	\right\}\right]
\end{aligned}
\end{equation}
which results in
\begin{equation}
\label{eq:kernel01}
	K(x,x{'};\epsilon)=
	\sqrt{\frac{m}{\ 2\pi\hbar\epsilon\ }\,}
	\exp\left[-\frac{m}{\ 2\hbar\epsilon\ }f{'}(x)f{'}(x{'})
	\left\{Q(x)-Q(x{'})\right\}^{2}-
	\frac{\ \epsilon\ }{\ \hbar\ }V_{\mathrm{eff}}(x,x{'})
	\right].
\end{equation}
We thus obtain a time sliced path integral
\begin{equation}
\label{eq:kernel02}
\begin{aligned}
	&\langle x_{\mathrm{F}}\vert e^{-\beta H/\hbar}\vert x_{\mathrm{I}}\rangle=
	\lim\limits_{N\to\infty}
	\left(\frac{m}{\ 2\pi\hbar\epsilon\ }\right)^{N/2}
	\int_{a}^{b}\prod_{i=1}^{N-1}dx_{i}\\
	&\hphantom{\langle x_{\mathrm{F}}\vert e^{-\beta H/\hbar}\vert x_{\mathrm{I}}\rangle=}\times
	\exp\left[-\sum_{j=1}^{N}\left\{\frac{m}{\ 2\hbar\epsilon\ }
	f{'}(x_{j})f{'}(x_{j-1})
	\left\{Q(x_{j})-Q(x_{j-1})\right\}^{2}+
	\frac{\ \epsilon\ }{\ \hbar\ }V_{\mathrm{eff}}(x_{j},x_{j-1})\right\}\right]
\end{aligned}
\end{equation}
where we have set $x_{\mathrm{F}}=x_{N}$ and $x_{\mathrm{I}}=x_{0}$.

We will observe in the following, if $x$ ranges over the whole real axis so that the equation $Q(x)-Q(x{'})=0$ has only a single solution $x-x{'}=0$, that the time sliced path integral given by \eqref{eq:kernel02} can be converted into
\begin{equation}
\label{eq:kernel03a}
\begin{aligned}
	&\langle x_{\mathrm{F}}\vert e^{-\beta H/\hbar}\vert x_{\mathrm{I}}\rangle\\
	=&
	\lim\limits_{N\to\infty}
	\left(\frac{m}{\ 2\pi\hbar\epsilon\ }\right)^{N/2}
	\int_{-\infty}^{\infty}\prod_{i=1}^{N-1}dx_{i}
	\exp\left[-\sum_{j=1}^{N}\left\{\frac{m}{\ 2\hbar\epsilon\ }
	(\Delta x_{j})^{2}+
	\frac{\ \epsilon\ }{\ \hbar\ }V(x_{j},x_{j-1})\right\}\right],
\end{aligned}
\end{equation}
where $\Delta x_{j}\equiv x_{j}-x_{j-1}$ and $V(x_{j},x_{j-1})=\langle x_{j}\vert V(x)\vert x_{j-1}\rangle$.
Obviously this is nothing but a time sliced path integral we usually obtain by making use of the completeness of the eigenvector of the momentum operator $p$.
It is, therefore, trivial result for systems defined on the whole real line and we are just making point transformation between systems defined on the whole real axis.
The assumption that the path integral is dominated by the contribution from the saddle point at $\Delta x_{j}=0$ will not be, however, fulfilled when both $a$ and $b$ are, or at least one of them is, finite. For such cases we have to extend the domain of the integration form $a<x<b$ to the whole real line in order to take contributions from multiple saddle points into account. This will be explained in the next section through solving examples explicitly and in an exact manner.

Let us check the validity of the short time kernel
\eqref{eq:kernel01} and observe that \eqref{eq:kernel02} is equivalent to \eqref{eq:kernel03a} which possesses an Euclidean Lagrangian
\begin{equation}
\label{eq:lagrange01}
  L_{\mathrm{E}}=\frac{m}{\ 2m\ }\dot{x}^{2}+V(X)
 \end{equation}
 in its exponent. To this aim we first consider, by assuming $a=-\infty$ and $b=\infty$,
\begin{equation}
\label{eq:check01}
	\psi(x,\beta+\epsilon)=\int_{-\infty}^{\infty}\!\!
	K(x,x{'};\epsilon)
	\psi(x{'},\beta)\,dx{'}
\end{equation}
for infinitesimally small $\epsilon$. 
By setting $x{'}=x+\eta$, we find
\begin{equation}
\begin{aligned}
	&f{'}(x)f{'}(x{'})
	\left\{Q(x)-Q(x{'})\right\}^{2}=
	\eta^{2}+\left\{f{'}(x)Q{''}(x)+
	\frac{\ f{''}(x)\ }{\ f{'}(x)\ }\right\}\eta^{3}\\
	&
	\hphantom{f{'}(x)f{'}(x{'})}
	+\left[\frac{1}{\ 4\ }\left\{f{'}(x)Q{''}(x)\right\}^{2}+
	\frac{1}{\ 3\ }f{'}(x)Q{'''}(x)+
	f{''}(x)Q{''}(x)+\frac{\ f{'''}(x)\ }{\ 2f{'}(x)\ }\right]\eta^{4}+
	O(\eta^{5})
\end{aligned}
\end{equation}
where use has been made of Taylor expansions for $f{'}(x{'})$ and $Q(x{'})$.
If we recall the definition of $Q(x)$, we observe that coefficients in the right hand side above are given by
\begin{equation}
	f{'}(x)Q{''}(x)+
	\frac{\ f{''}(x)\ }{\ f{'}(x)\ }=0
\end{equation}
and
\begin{equation}
	\frac{1}{\ 4\ }\left\{f{'}(x)Q{''}(x)\right\}^{2}+
	\frac{1}{\ 3\ }f{'}(x)Q{'''}(x)+
	f{''}(x)Q{''}(x)+\frac{\ f{'''}(x)\ }{\ 2f{'}(x)\ }=
	-\frac{1}{\ 12\ }\left[\left\{\frac{\ f{''}(x)\ }{\ f{'}(x)\ }\right\}^{2}-
	\frac{\ 2f{'''}(x)\ }{\ f{'}(x)\ }\right].
\end{equation}
In the same way, we expand $\psi(x{'},\beta)$ into the Taylor series to obtain
\begin{equation}
	\psi(x{'},\beta)=\psi(x,\beta)+
	\frac{\ \partial\psi(x,\beta)\ }{\ \partial x\ }\eta+
	\frac{1}{\ 2\ }\frac{\ \partial^{2}\psi(x,\beta)\ }{\ \partial x^{2}\ }
	\eta^{2}+O(\eta^{3}).
\end{equation}

The integration in the right hand side of \eqref{eq:check01} is now rewritten as
\begin{multline}
	\sqrt{\frac{m}{\ 2\pi\hbar\epsilon\ }\,}
	\int_{-\infty}^{\infty}\!\!\exp\left(-\frac{m}{\ 2\hbar\epsilon\ }\eta^{2}\right)
	\left\{1+\frac{m}{\ 24\hbar\epsilon\ }
	\left[\left\{\frac{\ f{''}(x)\ }{\ f{'}(x)\ }\right\}^{2}-
	\frac{\ 2f{'''}(x)\ }{\ f{'}(x)\ }\right]\eta^{4}\right\}\\
	\times\left\{1-\frac{\ \epsilon\ }{\ \hbar\ }V_{\mathrm{eff}}(x,x)\right\}
	\left\{\psi(x,\beta)+
	\frac{\ \partial\psi(x,\beta)\ }{\ \partial x\ }\eta+
	\frac{1}{\ 2\ }\frac{\ \partial^{2}\psi(x,\beta)\ }{\ \partial x^{2}\ }
	\eta^{2}\right\}
	\,d\eta
\end{multline}
by discarding irrelevant terms in the limit $\epsilon\to0$.
We thus obtain
\begin{equation}
\label{eq:psi_epsilon}
	\psi(x,\beta+\epsilon)=
	\psi(x,\beta)-\frac{\ \epsilon\ }{\ \hbar\ }\left\{
	-\frac{\ \hbar^{2}\ }{\ 2m\ }\frac{\partial^{2}\ }{\ \partial x^{2}\ }+
	V_{\mathrm{eff}}(x,x)-
	\frac{\ \hbar^{2}\ }{\ 8m\ }
	\left[\left\{\frac{\ f{''}(x)\ }{\ f{'}(x)\ }\right\}^{2}-
	\frac{\ 2f{'''}(x)\ }{\ f{'}(x)\ }\right]\right\}\psi(x,\beta).
\end{equation}
Since $V_{\mathrm{eff}}(x,x)$ is given by
\begin{equation}
	V_{\mathrm{eff}}(x,x)=\frac{\ \hbar^{2}\ }{\ 8m\ }
	\left[\left\{\frac{\ f{''}(x)\ }{\ f{'}(x)\ }\right\}^{2}-
	\frac{\ 2f{'''}(x)\ }{\ f{'}(x)\ }\right]+
	V(x),
\end{equation}
it is evident that the first term in $V_{\mathrm{eff}}(x,x)$ is precisely canceled by the last term in the right hand side of \eqref{eq:psi_epsilon}.
We hence observe that the result obtained above is equivalent to
\begin{equation}
	-\hbar\frac{\partial\ }{\ \partial\beta\ }\psi(x,\beta)=
	\left\{
	-\frac{\ \hbar^{2}\ }{\ 2m\ }\frac{\partial^{2}\ }{\ \partial x^{2}\ }+V(x)
	\right\}\psi(x,\beta).
\end{equation}
This is precisely the imaginary time version of the Schr\"{o}dinger equation.
We have thus confirmed the validity of the short time kernel given by \eqref{eq:kernel01}. This fact suggests that we can convert the time sliced path integral \eqref{eq:kernel02} into \eqref{eq:kernel03a} at least for a system on the whole real axis.

If the system is defined on a finite interval or on the half-line, we may find multiple critical points to be taken into account in carrying out the integration \eqref{eq:check01}. For such cases, we may set $x{'}=x_{\mathrm{C}}^{(n)}+\eta$ around a critical point $x_{\mathrm{C}}^{(n)}$.
Then, by making a change of variables given by
\begin{equation}
	\xi\equiv\sqrt{\frac{m}{\ 2\hbar\epsilon\ }\,}\eta,
\end{equation}
we can regard the integration with respect to $\xi$ as a Gaussian integration from $-\infty$ to $\infty$ for small $\epsilon$ even for the case both $a$ and $b$ are finite. To do so, however, we must extend the definition of the short time kernel outside the original domain $a<x{'}<b$ because $Q(x)$ was defined only on this interval with the positivity of $f{'}(x)$. This will require a carefull treatment, in particular, on the phase of the kernel depending on the detail of the system as we shall see in the next section.

Returning to the system on the whole real axis again, we rewrite $f{'}(x)$ and $f{'}(x{'})$ by expanding them around $x^{(\alpha)}\equiv \bar{x}-\alpha\Delta x$($\bar{x}\equiv(x+x{'})/2$, $\Delta x\equiv x-x{'}$) as
\begin{equation}
  f{'}(x)=f{'}(x^{(\alpha)})+
  f{''}(x^{(\alpha)})\left(\frac{\ 1\ }{\ 2\ }+\alpha\right)\Delta x+
  \frac{\ 1\ }{\ 2\ }f{'''}(x^{(\alpha)})
  \left(\frac{\ 1\ }{\ 2\ }+\alpha\right)^{2}(\Delta x)^{2}+\cdots
 \end{equation}
and
\begin{equation}
  f{'}(x{'})=f{'}(x^{(\alpha)})+
  f{''}(x^{(\alpha)})\left(\frac{\ 1\ }{\ 2\ }-\alpha\right)\Delta x+
  \frac{\ 1\ }{\ 2\ }f{'''}(x^{(\alpha)})
  \left(\frac{\ 1\ }{\ 2\ }-\alpha\right)^{2}(\Delta x)^{2}+\cdots
 \end{equation}
to find
\begin{equation}
\begin{aligned}
  &f{'}(x)f{'}(x{'})=\left\{f{'}(x^{(\alpha)})\right\}^{2}\\
  &\times\left[
	\vphantom{
	\left(\frac{\ f{''}(x^{(\alpha)})\ }{\ f{'}(x^{(\alpha)})\ }\right)^{2}}
	1+\alpha\frac{\ f{''}(x^{(\alpha)})\ }{\ f{'}(x^{(\alpha)})\ }\Delta x+
    \frac{\ 1\ }{\ 2\ }\left\{
	(\alpha^{2}+1/4)\frac{\ f{'''}(x^{(\alpha)})\ }{\ f{'}(x^{(\alpha)})\ }-
	\frac{\ 1\ }{\ 4\ }
	\left(\frac{\ f{''}(x^{(\alpha)})\ }{\ f{'}(x^{(\alpha)})\ }\right)^{2}
	\right\}(\Delta x)^{2}+\cdots
	\right]^{2}.
\end{aligned}
\end{equation}
Then, in the same way, we expand $Q(x)-Q(x{'})$ into a series as
\begin{equation}
\begin{aligned}
  &Q(x)-Q(x{'})\\
  =&\frac{\ \Delta x\ }{\ f{'}(x^{(\alpha)})\ }
  \left[
	\vphantom{
	\left(\frac{\ f{''}(x^{(\alpha)})\ }{\ f{'}(x^{(\alpha)})\ }\right)^{2}}
	1-\alpha\frac{\ f{''}(x^{(\alpha)})\ }{\ f{'}(x^{(\alpha)})\ }\Delta x+
    \left\{
	\left(\frac{\ f{''}(x^{(\alpha)})\ }{\ f{'}(x^{(\alpha)})\ }\right)^{2}-
	\frac{\ 2f{'''}(x^{(\alpha)})\ }{\ f{'}(x^{(\alpha)})\ }
	\right\}(\alpha^{2}+1/12)(\Delta x)^{2}+\cdots\right],
\end{aligned}
\end{equation}
so that we obtain
\begin{equation}
f{'}(x)f{'}(x{'})\{Q(x)-Q(x{'})\}^{2}=
	\{R(x^{(\alpha)},\Delta x)\Delta x\}^{2}
\end{equation}
in which $R(x^{(\alpha)},\Delta x)$ being given by
\begin{equation}
\label{eq:R}
	R(x^{(\alpha)},\Delta x)
	=1-\frac{\ 1\ }{\ 24\ }
	\left\{
	\left(\frac{\ f{''}(x^{(\alpha)})\ }{\ f{'}(x^{(\alpha)})\ }\right)^{2}-
	\frac{\ 2f{'''}(x^{(\alpha)})\ }{\ f{'}(x^{(\alpha)})\ }
	\right\}(\Delta x)^{2}+\cdots.
\end{equation}

According to ref.\onlinecite{Sakoda2017}, a time sliced path integral with a complicated kinetic term given by
\begin{equation}
\label{eq:kinetic}
\frac{\ 1\ }{\ 2\lambda\ }\{\Delta x_{j}R(x^{(\alpha)}_{j},\Delta x_{j})\}^{2},
\quad
R(x^{(\alpha)}_{j},\Delta x_{j})=1+a_{2}(x^{(\alpha)}_{j})\Delta x_{j}+
a_{3}(x^{(\alpha)}_{j})(\Delta x_{j})^{2}+\cdots,
\end{equation}
can be converted into its equivalent with the standard kinetic term:
\begin{equation}
\label{eq:ss2017}
	\begin{aligned}
	&\left(\frac{\ 1\ }{\ 2\pi\lambda\ }\right)^{N/2}\!\!\!\!\!
\int_{-\infty}^{\infty}\!\prod_{i=1}^{N-1}dx_{i}\,
\exp\left[-\sum_{j=1}^{N}
\frac{\ 1\ }{\ 2\lambda\ }\{\Delta x_{j}R(x^{(\alpha)}_{j},\Delta x_{j})\}^{2}
\right]\\
=&
\left(\frac{\ 1\ }{\ 2\pi\lambda\ }\right)^{N/2}\!\!\!\!\!
\int_{-\infty}^{\infty}\!\prod_{i=1}^{N-1}dx_{i}\,
\exp\left[-\sum_{j=1}^{N}\left\{\frac{\ (\Delta x_{j})^{2}\ }{\ 2\lambda\ }
+\lambda U_{\mathrm{add}}(x^{(\alpha)}_{j})
\right\}\right],
\end{aligned}
\end{equation}
where $\lambda=\hbar\epsilon/m$ and the additional potential being given by
\begin{equation}
\label{eq:Uadd}
	U_{\mathrm{add}}(x)=3a_{3}(x)-3(\alpha-1/2)a_{2}{'}(x)-2\{a_{2}(x)\}^{2}
\end{equation}
which is equivalent to
\begin{equation}
\label{eq:Vadd}
	V_{\mathrm{add}}(x)=\frac{\ \hbar^{2}\ }{\ m\ }\left\{
	3a_{3}(x)-3(\alpha-1/2)a_{2}{'}(x)-2\{a_{2}(x)\}^{2}\right\}.
\end{equation}
Here we have written $\lambda U_{\mathrm{add}}(x)$ as
\begin{equation}
	\lambda U_{\mathrm{add}}(x)=\frac{\ \epsilon\ }{\ \hbar\ }
	V_{\mathrm{add}}(x).
\end{equation}
In view of the series expansion \eqref{eq:R} above, we immediately obtain an additional potential
\begin{equation}
  V_{\mathrm{add}}(x)=-\frac{\ \hbar^{2}\ }{\ 8m\ }\left\{
	\left(\frac{\ f{''}(x)\ }{\ f{'}(x)\ }\right)^{2}-
	\frac{\ 2f{'''}(x)\ }{\ f{'}(x)\ }
	\right\}
\end{equation}
as contribution to the path integral through rewriting \eqref{eq:kernel01} to have the kinetic term in the standard form. Since the effective potential in the exponent of \eqref{eq:kernel02} can be taken to be
\begin{equation}
\label{eq:V_eff}
	V_{\mathrm{eff}}(x_{j},x_{j-1})=\frac{\hbar^{2}}{\ 8m\ }\left[\left\{
	\frac{\ f{''}(x^{(\alpha)}_{j})\ }{\ f{'}(x^{(\alpha)}_{j})\ }\right\}^{2}-
	\frac{\ 2f{'''}(x^{(\alpha)}_{j})\ }{\ f{'}(x^{(\alpha)}_{j})\ }\right]+
	V(x^{(\alpha)}_{j}),
\end{equation}
the cancellation of the term, that emerged by rewriting the Hamiltonian in terms of $P$, by $V_{\mathrm{add}}(x^{(\alpha)}_{j})$ is evident. We therefore obtain the time sliced path integral \eqref{eq:kernel03a} as an equivalent one for \eqref{eq:kernel02}. Here, a comment is in need; we have employed the $\alpha$-ordering for evaluating the potential term of the path integral at $x^{(\alpha)}$. This is useful to check the ordering independence of the path integral for the case $x$ takes its value from $-\infty$ to $\infty$. For systems defined on the half-line or on a finite interval, there will be other natural and suitable scheme for evaluating potential term. For example, as the midpoint prescription for a system on the half-line, we may write $V(x,x{'})$ in the path integral as $V(\sqrt{xx{'}\,})$ because the geometric mean will be more suitable than taking the arithmetic average as $V(\bar{x})$.

\section{Systems with non-trivial geometry}
\label{sec:examples}
In this section, we consider path integrals for systems confined within a finite interval and for systems on the half-line as examples for the time sliced path integral \eqref{eq:kernel02}. For a system defined on $0<x<L$, we may expect that $f{'}(x)$ behaves like $f{'}(x)\propto x$ near the origin and like $f{'}(x)\propto L-x$ when $x$ approaches to $L$ so that the new coordinate $Q(x)$ takes its value from $-\infty$ to $\infty$. We may further assume that the Hamiltonian should have a term proportional to $1/\{f{'}(x)\}^{2}$ as a potential for the physical requirement that wave functions must vanish at boundaries. In the same way, we may assume $f{'}(x)\propto x$ near the origin and a potential proportional to $1/x^{2}$ for a system on the half-line. If we introduce potential terms in this way, we can expect wave functions to behave like $\{f{'}(x)\}^{\alpha}$ where $\alpha$ being determined by the coefficient of the potential term. If $\alpha$ is not an integer, this behavior makes wave functions be multi-valued when considered on the extended domain which is needed to take effects of reflections at boundaries into account. Since $f{'}(x)$ vanishes at the boundary, the sign of this function will change if we put $x$ outside the original domain. It is, therefore, important to take the phase factor of the short time kernel into account when we extend the domain to evaluate contributions from multiple saddle points.  

\subsection{Systems on a finite interval}
We consider here a system described by the Hamiltonian
\begin{equation}
\label{eq:PTham}
  H=\frac{1}{\ 2m\ }p^{2}+
  \frac{\ \nu(\nu-1){\mathcal R}\ }{\ \sin^{2}\theta\ },\quad
  {\mathcal R}\equiv\frac{\hbar^{2}}{\ 2ma^{2}\ },\
  \theta\equiv\frac{\ x\ }{\ a\ },\
  a\equiv\frac{\ L\ }{\ \pi\ },
\end{equation}
in which we assume that $\nu\ge1/2$, on a finite interval $0<x<L$.
A suitable choice for the function $f(x)$ will be given by
\begin{equation}
	f(x)=-a\cos\theta
\end{equation}
to yield
\begin{equation}
	Q(x)=a\log\{\tan(\theta/2)\}
\end{equation}
and
\begin{equation}
	P=-\frac{\ i\hbar\ }{\ 2a\ }
	\left(\sin\theta\frac{\ d\ \ }{\ d\theta\ }+
	\frac{\ d\ \ }{\ d\theta\ }\sin\theta\right).
\end{equation}
The eigenfunction of this operator is given by
\begin{equation}
	\psi_{P{'}}(x)=\frac{1}{\ \sqrt{2\pi\hbar\sin\theta\,}\ }
	\exp\left[\frac{\ i\ }{\ \hbar\ }aP{'}\log\{\tan(\theta/2)\}\right]
\end{equation}
whose completeness can be seen as, $\theta{'}$ corresponding to $x{'}$ like $\theta$ for $x$ above,
\begin{equation}
\label{eq:delta01}
	\int_{-\infty}^{\infty}\!\!\psi_{P}(x)\psi^{*}_{P}(x{'})\,dP=
	\frac{1}{\ a\sqrt{\sin\theta\sin\theta{'}\,}\ }
	\delta(\log\{\tan(\theta/2)\}-\log\{\tan(\theta{'}/2)\})=
	\delta(x-x{'})
\end{equation}
for $x$ and $x{'}$ belonging to the interval $(0,L)$.
We can utilize this completeness relation to find the short time kernel
\begin{equation}
\label{eq:kernel04}
	{\mathcal K}(\theta,\theta{'};\epsilon)=\frac{1}{\ \sqrt{2\pi\lambda\,}\ }
	\exp\left[-\frac{1}{\ 2\lambda\ }
	\sin\theta\sin\theta{'}
	\left\{\log\frac{\ \tan(\theta/2)\ }{\ \tan(\theta{'}/2)\ }
	\right\}^{2}-
	\lambda U_{\mathrm{eff}}(\theta,\theta{'})
	\right],\quad
	\lambda\equiv\frac{\ \hbar\epsilon\ }{\ ma^{2}\ },
\end{equation}
where the effective potential being given by
\begin{equation}
	U_{\mathrm{eff}}(\theta,\theta{'})=\frac{\ 1\ }{\ 8\ }
	\left(1+\frac{1}{\ \sin\theta\sin\theta{'}\ }\right)+
	\frac{\ 1\ }{\ 2\ }\frac{\nu(\nu-1)}{\ \sin\theta\sin\theta{'}\ }
\end{equation}
for the Hamiltonian given by \eqref{eq:PTham}. Note that the kernel \eqref{eq:kernel04} is normalized to fit the integration with respect to $\theta$ instead of $x$. This allows us to make expressions below considerably simple. If we devide it by $a$, we will obtain the one corresponding to the integration with respect to $x$. It will be clear that we have employed the geometric mean as the ordering prescription for $U_{\mathrm{eff}}(\theta,\theta{'})$.

In addition to the saddle point located at $\theta-\theta{'}=0$, there may be contributions from paths reflected at boundaries. To take into account of these contributions, we need to extend the domain of variables outside the original one. By keeping $\theta{'}$ inside the original domain, we here consider the behavior of the kernel as a function of $\theta$ on the extended domain. Let us first study the case $\pi<\theta<2\pi$. We must replace $\sin\theta$ and $\tan(\theta/2)$ by their absolute values in the kernel \eqref{eq:kernel04}. In addition to this, we have to determine the phase factor which the kernel acquires when $\theta$ goes outside the original domain. This can be achieved by considering solutions of the stationary Schr\"{o}dinger equation for the Hamiltonian \eqref{eq:PTham}. In view of singularities of the potential at boundaries, we see that eigenfunctions must be proportional to $(\sin\theta)^{\nu}$ if $\nu\ge1/2$ as assumed above\cite{S.Ohya}.
Taking into account of this behavior and of the fact that the kernel can be expressed as the eigenfunction expansion, we see that the kernel will acquire the phase $\nu\arg(\sin\theta)$ when $\theta$ goes from the original domain to $\pi<\theta<2\pi$. On the original region, we choose such that $\arg(\sin\theta)=0$ and $\arg(\sin\theta)=\pi$ on $\pi<\theta<2\pi$. This is equivalent to circumvent the singularity at $\theta=\pi$ by going along an infinitesimally small hemicircle below $\pi$. We thus obtain
\begin{equation}
\label{eq:kernel04a}
	{\mathcal K}^{(1)}(\theta,\theta{'};\epsilon)=
	\frac{e^{\nu\pi i}}{\ \sqrt{2\pi\lambda\,}\ }
	\exp\left[-\frac{1}{\ 2\lambda\ }
	\abs{\sin\theta}\sin\theta{'}
	\left\{\log\frac{\ \abs{\tan(\theta/2)}\ }{\ \tan(\theta{'}/2)\ }
	\right\}^{2}-
	\lambda U_{\mathrm{eff}}(\theta,\theta{'})
	\right]
\end{equation}
as the short time kernel for $\pi<\theta<2\pi$. Here the potential $U_{\mathrm{eff}}(\theta,\theta{'})$ must be also expressed by $\abs{\sin\theta}$ of course.
In the same way, we define $\arg(\sin\theta)$ such that $\arg(\sin\theta)=k\pi$ for $k\pi<\theta<(k+1)\pi$ ($n=0,\,\pm1,\,\pm2,\,\dots$) to obtain
\begin{equation}
\label{eq:kernel04b}
	{\mathcal K}^{(k)}(\theta,\theta{'};\epsilon)=
	\frac{e^{k\nu\pi i}}{\ \sqrt{2\pi\lambda\,}\ }
	\exp\left[-\frac{1}{\ 2\lambda\ }
	\abs{\sin\theta}\sin\theta{'}
	\left\{\log\frac{\ \abs{\tan(\theta/2)}\ }{\ \tan(\theta{'}/2)\ }
	\right\}^{2}-
	\lambda U_{\mathrm{eff}}(\theta,\theta{'})
	\right]
\end{equation}
as the short time kernel for $\theta$ in $k$-th domain.
It will be evident that ${\mathcal K}^{(k)}(\theta,\theta{'};\epsilon)$ possesses a saddle point at $\theta=\theta{'}+k\pi$ if $k$ is an even integer. If $k$ is an odd integer, the saddle point is found at $\theta=-\theta{'}+(k+1)\pi$.
Hence contributions from paths reflected even times at boundaries yields
\begin{equation}
\label{eq:k_eve01}
	{\mathcal K}^{(\mathrm{e})}(\theta,\theta{'};\epsilon)
	=\sum_{k=-\infty}^{\infty}
	\frac{\ e^{2k\nu\pi i}\ }{\ \sqrt{2\pi\lambda\,}\ }
	\exp\left[-\frac{1}{\ 2\lambda\ }(\theta-\theta{'}-2k\pi)^{2}-
	\frac{\ \lambda\ }{\ 2\ }
	\frac{\nu(\nu-1)}{\ \sin\theta\sin\theta{'}\ }
	\right]
\end{equation}
while contributions from paths reflected even times at boundaries being given by
\begin{equation}
\label{eq:k_odd01}
	{\mathcal K}^{(\mathrm{o})}(\theta,\theta{'};\epsilon)
	=\sum_{k=-\infty}^{\infty}
	\frac{\ e^{(2k-1)\nu\pi i}\ }{\ \sqrt{2\pi\lambda\,}\ }
	\exp\left[-\frac{1}{\ 2\lambda\ }(\theta+\theta{'}-2k\pi)^{2}+
	\frac{\ \lambda\ }{\ 2\ }
	\frac{\nu(\nu-1)}{\ \sin\theta\sin\theta{'}\ }
	\right].
\end{equation}
Here use has been made of the technique of ref.\onlinecite{Sakoda2017} to convert the complicated kinetic term in short time kernels.
Note that $\arg(\sin\theta)=2k\pi$ for each term in the sum in \eqref{eq:k_eve01} and $\arg(\sin\theta)=(2k-1)\pi$ in \eqref{eq:k_odd01}.
Keeping these in mind, we consider the asymptotic form of the modified Bessel function $I_{\nu-1/2}(\sin\theta\sin\theta{'}/\lambda)$ for infinitesimally small $\lambda$. On the original domain, this function has the asymptotic expression
\begin{equation}
\label{eq:modBessel01}
	I_{\nu-1/2}\left(\frac{\ \sin\theta\sin\theta{'}\ }{\ \lambda\ }\right)\sim
	\sqrt{\frac{\lambda}{\ 2\pi\sin\theta\sin\theta{'}\ }\,}
	\exp\left\{\frac{\ \sin\theta\sin\theta{'}}{\lambda}-
	\frac{\ \lambda\ }{\ 2\ }
	\frac{\ \nu(\nu-1)\ }{\ \sin\theta\sin\theta{'}\ }\right\}
\end{equation}
and if $\arg(\sin\theta)=2k\pi$, by analytic continuation, it changes to
\begin{equation}
\label{eq:modBessel01a}
	e^{2k\nu\pi i}
	\sqrt{\frac{\lambda}{\ 2\pi\sin\theta\sin\theta{'}\ }\,}
	\exp\left\{\frac{\ \sin\theta\sin\theta{'}}{\lambda}-
	\frac{\ \lambda\ }{\ 2\ }
	\frac{\ \nu(\nu-1)\ }{\ \sin\theta\sin\theta{'}\ }\right\}.
\end{equation}
On the other hand, if $\arg(\sin\theta)=(2k-1)\pi$, the asymptotic form reads
\begin{equation}
\label{eq:modBessel01b}
	e^{(2k-1)\nu\pi i}
	\sqrt{\frac{\lambda}{\ 2\pi\abs{\sin\theta}\sin\theta{'}\ }\,}
	\exp\left\{-\frac{\ \sin\theta\sin\theta{'}}{\lambda}+
	\frac{\ \lambda\ }{\ 2\ }
	\frac{\ \nu(\nu-1)\ }{\ \sin\theta\sin\theta{'}\ }\right\}.
\end{equation}

Remembering the periodicity and its Taylor expansion of $1-\cos(\theta-\theta{'})$, we can replace $(\theta-\theta{'}-2k\pi)^{2}/2$ in \eqref{eq:k_eve01} by $1-\cos(\theta-\theta{'})$ with an additional constant potential term\cite{Sakoda2017}. Then, by comparing with the asymptotic form given by \eqref{eq:modBessel01a}, we observe that the sum in \eqref{eq:k_eve01} is nothing but the contributions from saddle points of
\begin{equation}
\label{eq:kernel05}
	{\mathcal K}(\theta,\theta{'};\epsilon)=
	\exp\left(-\frac{\ 1-\cos\theta\cos\theta{'}\ }{\lambda}-
	\frac{\ \lambda\ }{\ 8\ }\right)
	\frac{\ (\sin\theta\sin\theta{'})^{1/2}\ }{\lambda}
	I_{\nu-1/2}\left(\frac{\ \sin\theta\sin\theta{'}}{\lambda}\right)
\end{equation}
located at $\theta=\theta{'}+2k\pi$ ($k=0,\,\pm1,\,\pm2,\,\dots$).
In the same way, we find that the sum in \eqref{eq:k_odd01} can be regarded as contributions from saddle points at $\theta=-\theta{'}+2k\pi$ ($k=0,\,\pm1,\,\pm2,\,\dots$) for the same expression \eqref{eq:kernel05}.
Therefore ${\mathcal K}(\theta,\theta{'};\epsilon)$ above involves all contributions from paths reflected at boundaries and we can regard it as the short time kernel for infinitesimally small $\lambda$. In this regard, the expression
\begin{equation}
	{\mathcal K}(\theta,\theta{'};\epsilon)
	=
	{\mathcal K}^{(\mathrm{e})}(\theta,\theta{'};\epsilon)
	+
	{\mathcal K}^{(\mathrm{o})}(\theta,\theta{'};\epsilon)
\end{equation}
can be regarded as the decomposition of the Feynman kernel into the sum over paths\cite{Sakoda2018b}.

In order to deduce eigenfunctions of the Hamiltonian from the short time kernel \eqref{eq:kernel05}, we here make use of a formula(see e.g. Ch. 11.5 of ref.\onlinecite{Watson} or Ch. 8.8 of ref.\onlinecite{KleinertBook})
\begin{equation}
\label{eq:bessel_formula}
	\begin{aligned}
	&\frac{\ (\sin\theta\sin\theta{'})^{1/2}\ }{\lambda}
	\exp\left({\frac{\ \cos\theta\cos\theta{'}\ }{\lambda}}\right)
	I_{\nu-1/2}\left(\frac{\ \sin\theta\sin\theta{'}}{\lambda}\right)\\
	=&
	\frac{\ 2^{2\nu}\{\varGamma(\nu)\}^{2}\ }{\ \sqrt{2\pi\lambda\,}\ }
	(\sin\theta\sin\theta{'})^{\nu}
	\sum_{n=0}^{\infty}\frac{\ n!(\nu+n)\ }{\ \varGamma(2\nu+n)\ }
	I_{\nu+n}\left(\frac{\ 1\ }{\ \lambda\ }\right)
	C_{n}^{\nu}(\cos\theta)C_{n}^{\nu}(\cos\theta{'}).
\end{aligned}
\end{equation}
Comparing the left hand side above with the right hand side of \eqref{eq:kernel05}, we find
\begin{equation}
\label{eq:kernel_SPT01}
	{\mathcal K}(\theta,\theta{'};\epsilon)=
	\sum_{n=0}^{\infty}\left\{\sqrt{\frac{\ 2\pi\ }{\lambda}\,}
	I_{\nu+n}\left(\frac{\ 1\ }{\ \lambda\ }\right)
	\exp\left(-\frac{\ 1\ }{\ \lambda\ }-\frac{\ \lambda\ }{\ 8\ }\right)
	\right\}
	\phi_{n}^{(\nu)}(\theta)\phi_{n}^{(\nu)}(\theta{'}),
\end{equation}
where the eigenfunction $\phi_{n}^{(\nu)}(\theta)$ being given by
\begin{equation}
\label{eq:eigenfunction01}
	\phi_{n}^{(\nu)}(\theta)=
	2^{\nu}\varGamma(\nu)
	\sqrt{\frac{\ n!(\nu+n)\ }{\ 2\pi\varGamma(2\nu+n)\ }\,}
	(\sin\theta)^{\nu}C_{n}^{\nu}(\cos\theta)
\end{equation}
in terms the Gegenbauer polynomial $C_{n}^{\nu}(\cos\theta)$.
If we make use of the orthogonality
\begin{equation}
	\int_{0}^{\pi}\!\!\phi_{n}^{(\nu)}(\theta)
	\phi_{n{'}}^{(\nu)}(\theta)\,d\theta=\delta_{n\,n{'}},
\end{equation}
we can easily check that there holds
\begin{equation}
\label{eq:reproducing}
	\int_{0}^{\pi}{\mathcal K}(\theta,\theta{'};\epsilon)
	{\mathcal K}(\theta{'},\theta{''};\epsilon)\,d\theta{'}=
	\sum_{n=0}^{\infty}\left\{\sqrt{\frac{\ 2\pi\ }{\lambda}\,}
	I_{\nu+n}\left(\frac{\ 1\ }{\ \lambda\ }\right)
	\exp\left(-\frac{\ 1\ }{\ \lambda\ }-\frac{\ \lambda\ }{\ 8\ }\right)
	\right\}^{2}
	\phi_{n}^{(\nu)}(\theta)\phi_{n}^{(\nu)}(\theta{'}).
\end{equation}
This can be repeated $N-1$ times to result in
\begin{equation}
\begin{aligned}
\label{eq:kernel_SPT02}
	\langle\theta\vert e^{-\beta H/\hbar}\vert\theta{'}\rangle
	=&\lim\limits_{N\to\infty}
	\sum_{n=0}^{\infty}\left\{\sqrt{\frac{\ 2\pi\ }{\lambda}\,}
	I_{\nu+n}\left(\frac{\ 1\ }{\ \lambda\ }\right)
	\exp\left(-\frac{\ 1\ }{\ \lambda\ }-\frac{\ \lambda\ }{\ 8\ }\right)
	\right\}^{N}
	\phi_{n}^{(\nu)}(\theta)\phi_{n}^{(\nu)}(\theta{'})\\
	=&
	\sum_{n=0}^{\infty}
	e^{-\beta E_{n}^{(\nu)}/\hbar}
	\phi_{n}^{(\nu)}(\theta)\phi_{n}^{(\nu)}(\theta{'}),
\end{aligned}
\end{equation}
in which we have written $\vert\theta\rangle=\vert x\rangle\sqrt{a}$ and $E_{n}^{(\nu)}=(n+\nu)^{2}{\mathcal R}$($n=0,\,1,\,2,\,\dots$) for energy eigenvalues. This is the eigenfunction expansion of the Feynman kernel for a finite imaginary time $\beta$. Thus we have obtained eigenvalues and corresponding eigenfunctions for the Hamiltonian \eqref{eq:PTham} solely by means of path integral method.

We now generalize the consideration above to a system described by the Hamiltonian
\begin{equation}
\label{eq:PTham2}
  H=\frac{1}{\ 2m\ }p^{2}+
  \frac{\ {\mathcal R}\ }{\ 4\ }\left\{
  \frac{\mu(\mu-1)}{\ \cos^{2}(\theta/2)\ }+
  \frac{\nu(\nu-1)}{\ \sin^{2}(\theta/2)\ }
  \right\}
\end{equation}
on the same domain $0<x<L$.
The short time kernel for this system will be same as \eqref{eq:kernel04} if we replace the effective potential to the one given by
\begin{equation}
	U_{\mathrm{eff}}(\theta,\theta{'})=\frac{\ 1\ }{\ 8\ }
	\left(1+\frac{1}{\ \sin\theta\sin\theta{'}\ }\right)+
	\frac{\ 1\ }{\ 8\ }\left\{
	\frac{\mu(\mu-1)}{\ \cos(\theta/2)\cos(\theta{'}/2)\ }+
	\frac{\nu(\nu-1)}{\ \sin(\theta/2)\sin(\theta{'}/2)\ }\right\}.
\end{equation}
In view of the singularities of the potential, eigenfunctions of the Hamiltonian must be proportional to $\sin^{\nu}(\theta/2)\cos^{\mu}(\theta/2)$. This determines the phase factor of the kernel when extended to outside the original domain $0<\theta<\pi$. Again, by keeping $\theta{'}$ within the original domain, we consider the kernel as a functions of $\theta$.
In the original region, we set $\arg(\sin(\theta/2))=0$ as well as $\arg(\cos(\theta/2))=0$ and we define the way to avoid the singularity at $k\pi$ ($k=0,\,\pm1,\,\pm2,\,\dots$) to go along infinitesimally small hemicircle blow $k\pi$. Then, if $\theta$ belongs to $(2k-1)\pi<\theta<(2k+1)\pi$, the argument of $\cos(\theta/2)$ is given by $\arg(\cos(\theta/2))=k\pi$. In the same way, if $2k\pi<\theta<2(k+1)\pi$, we get $\arg(\sin(\theta/2))=k\pi$. The phase of the short time kernel will be then given by $\mu\arg(\cos(\theta/2))+\nu\arg(\sin(\theta/2))$.
Corresponding to the number of reflections at $\theta=0$ and $\theta=\pi$, we obtain four kinds of contributions to the short time kernel.
The first one is given by
\begin{equation}
\label{eq:PTkernel01}
\begin{aligned}
	&{\mathcal K}^{(\mathrm{e},\mathrm{e})}(\theta,\theta{'};\epsilon)
	=
	\sum_{k=-\infty}^{\infty}
	\frac{\ e^{2k(\mu+\nu)\pi i}\ }{\ \sqrt{2\pi\lambda\,}\ }\\
	&\times\exp\left[
	-\frac{1}{\ 2\lambda\ }(\theta-\theta{'}-4k\pi)^{2}-
	\frac{\ \lambda\ }{\ 8\ }\left\{
	\frac{\ \mu(\mu-1)\ }{\ \cos(\theta/2)\cos(\theta{'}/2)\ }+
	\frac{\ \nu(\nu-1)\ }{\ \sin(\theta/2)\sin(\theta{'}/2)\ }\right\}\right].
\end{aligned}
\end{equation}
Here the symbol $(\mathrm{e},\mathrm{e})$ above designates that this kernel is the sum of contributions from paths reflected even times at both boundaries. If we write $(\mathrm{e},\mathrm{o})$, this means those from paths reflected even times by the one at $\theta=0$ and odd times by the one at $\theta=\pi$.
Other contributions from reflected paths are given by
\begin{equation}
\label{eq:PTkernel_eo}
\begin{aligned}
	&{\mathcal K}^{(\mathrm{e},\mathrm{o})}(\theta,\theta{'};\epsilon)
	=
	\sum_{k=-\infty}^{\infty}
	\frac{\ e^{\{2k(\mu+\nu)+\mu\}\pi i}\ }{\ \sqrt{2\pi\lambda\,}\ }\\
	&\times\exp\left[
	-\frac{1}{\ 2\lambda\ }\left\{\theta+\theta{'}-(4k+2)\pi\right\}^{2}-
	\frac{\ \lambda\ }{\ 8\ }\left\{
	-\frac{\ \mu(\mu-1)\ }{\ \cos(\theta/2)\cos(\theta{'}/2)\ }+
	\frac{\ \nu(\nu-1)\ }{\ \sin(\theta/2)\sin(\theta{'}/2)\ }\right\}\right],
\end{aligned}
\end{equation}
\begin{equation}
\label{eq:PTkernel_oe}
\begin{aligned}
	&{\mathcal K}^{(\mathrm{o},\mathrm{e})}(\theta,\theta{'};\epsilon)
	=
	\sum_{k=-\infty}^{\infty}
	\frac{\ e^{\{(2k+1)(\mu+\nu)+\mu\}\pi i}\ }{\ \sqrt{2\pi\lambda\,}\ }\\
	&\times\exp\left[
	-\frac{1}{\ 2\lambda\ }\{\theta+\theta{'}-4(k+1)\pi\}^{2}-
	\frac{\ \lambda\ }{\ 8\ }\left\{
	\frac{\ \mu(\mu-1)\ }{\ \cos(\theta/2)\cos(\theta{'}/2)\ }
	-\frac{\ \nu(\nu-1)\ }{\ \sin(\theta/2)\sin(\theta{'}/2)\ }\right\}\right]
\end{aligned}
\end{equation}
and
\begin{equation}
\label{eq:PTkernel_oo}
\begin{aligned}
	&{\mathcal K}^{(\mathrm{o},\mathrm{o})}(\theta,\theta{'};\epsilon)
	=
	\sum_{k=-\infty}^{\infty}
	\frac{\ e^{(2k+1)(\mu+\nu)\pi i}\ }{\ \sqrt{2\pi\lambda\,}\ }\\
	&\times\exp\left[
	-\frac{1}{\ 2\lambda\ }\left\{\theta-\theta{'}-(4k+2)\pi\right\}^{2}-
	\frac{\ \lambda\ }{\ 8\ }\left\{
	-\frac{\ \mu(\mu-1)\ }{\ \cos(\theta/2)\cos(\theta{'}/2)\ }
	-\frac{\ \nu(\nu-1)\ }{\ \sin(\theta/2)\sin(\theta{'}/2)\ }\right\}\right].
\end{aligned}
\end{equation}
Taking into account of $\arg(\sin(\theta/2))$ and $\arg(\cos(\theta/2))$ in partial kernels above, we consider the asymptotic form of the product of modified Bessel functions
\[
	I_{\mu-1/2}\left(
	\frac{\ 4\cos(\theta/2)\cos(\theta{'}/2)\ }{\lambda}\right)
	I_{\nu-1/2}\left(
	\frac{\ 4\sin(\theta/2)\sin(\theta{'}/2)\ }{\lambda}\right)
\]
to recognize that these are contributions of saddle points of the kernel
\begin{equation}
\begin{aligned}
\label{eq:PTkernel04}
	{\mathcal K}(\theta,\theta{'};\epsilon)=&
	\frac{1}{\ \sqrt{2\pi\lambda\,}\ }
	\exp\left(-\frac{\ 4\ }{\ \lambda\ }-\frac{\ \lambda\ }{\ 32\ }\right)
	\frac{\ 8\pi\ }{\lambda}
	\left(
	\cos\frac{\ \theta\ }{\ 2\ }\cos\frac{\ \theta{'}\ }{\ 2\ }
	\right)^{1/2}
	\left(
	\sin\frac{\ \theta\ }{\ 2\ }\sin\frac{\ \theta{'}\ }{\ 2\ }
	\right)^{1/2}\\
	&\times
	I_{\mu-1/2}\left(\frac{\ 4\ }{\ \lambda\ }
	\cos\frac{\ \theta\ }{\ 2\ }\cos\frac{\ \theta{'}\ }{\ 2\ }\right)
	I_{\nu-1/2}\left(\frac{\ 4\ }{\ \lambda\ }
	\sin\frac{\ \theta\ }{\ 2\ }\sin\frac{\ \theta{'}\ }{\ 2\ }\right)	
\end{aligned}
\end{equation}
for infinitesimally small $\lambda$.
Therefore we can regard ${\mathcal K}(\theta,\theta{'};\epsilon)$ as the short time kernel for the system under consideration.

Deduction of eigenfunctions and eigenvalues of the Hamiltonian \eqref{eq:PTham2} is achieved if we make use of a formula,which is equivalent to Bateman's expansion(see e.g. Ch. 11.6 in ref.\onlinecite{Watson} or Ch. 8.8 in ref.\onlinecite{KleinertBook}), given by
\begin{equation}
\begin{aligned}
\label{eq:bateman}
	&\frac{\ 2\ }{\ \lambda\ }
	\left(
	\cos\frac{\ \theta\ }{\ 2\ }\cos\frac{\ \theta{'}\ }{\ 2\ }\right)^{1/2}
	\left(
	\sin\frac{\ \theta\ }{\ 2\ }\sin\frac{\ \theta{'}\ }{\ 2\ }\right)^{1/2}\\
	&\times
	I_{\mu-1/2}\left(\frac{\ 4\ }{\ \lambda\ }
	\cos\frac{\ \theta\ }{\ 2\ }\cos\frac{\ \theta{'}\ }{\ 2\ }\right)
	I_{\nu-1/2}\left(\frac{\ 4\ }{\ \lambda\ }
	\sin\frac{\ \theta\ }{\ 2\ }\sin\frac{\ \theta{'}\ }{\ 2\ }\right)\\
	=&
	\sum_{n=0}^{\infty}
	I_{\mu+\nu+2n+1}\left(\frac{\ 4\ }{\ \lambda\ }\right)
	\frac
	{\ (\mu+\nu+2n)n!\varGamma(\mu+\nu+n)\ }
	{\ \varGamma(\mu+n+1/2)\varGamma(\nu+n+1/2)\ }\\
	&
	\times
	P_{n}^{(\nu-1/2,\mu-1/2)}(\cos\theta)
	P_{n}^{(\nu-1/2,\mu-1/2)}(\cos\theta{'})
	\left(
	\cos\frac{\ \theta\ }{\ 2\ }\cos\frac{\ \theta{'}\ }{\ 2\ }\right)^{\mu}
	\left(
	\sin\frac{\ \theta\ }{\ 2\ }\sin\frac{\ \theta{'}\ }{\ 2\ }\right)^{\nu}
\end{aligned}
\end{equation}
where $P_{n}^{(\nu-1/2,\mu-1/2)}(\cos\theta)=(\varGamma(\nu+n+1/2)/(n!\varGamma(\nu+1/2)){}_{2}F_{1}(-n,\mu+\nu+n;\nu+1/2;\sin^{2}\theta)$ is the Jacobi polynomial.
By comparing the left hand side above with \eqref{eq:PTkernel04}, we find that the short time kernel can be rewritten as
\begin{equation}
\label{eq:PTkernel05}
	{\mathcal K}(\theta,\theta{'};\epsilon)=
	\sum_{n=0}^{\infty}\left\{\sqrt{\frac{\ 8\pi\ }{\lambda}\,}
	I_{\mu+\nu+2n+1}\left(\frac{\ 4\ }{\ \lambda\ }\right)
	\exp\left(-\frac{\ 4\ }{\ \lambda\ }-\frac{\ \lambda\ }{\ 32\ }\right)
	\right\}
	\phi_{n}^{(\mu,\nu)}(\theta)\phi_{n}^{(\mu,\nu)}(\theta{'}),
\end{equation}
where the eigenfunction belonging to the eigenvalue $E_{n}^{(\mu,\nu)}=\{n+(\mu+\nu)/2\}^{2}{\mathcal R}$ is given by
\begin{equation}
\label{eq:eigenfunctionPT}
	\phi_{n}^{(\mu,\nu)}(\theta)=
	\sqrt{\frac{\ (\mu+\nu+2n)n!\varGamma(\mu+\nu+n)\ }
	{\ \varGamma(\mu+n+1/2)\varGamma(\nu+n+1/2)\ }\,}
	P_{n}^{(\nu-1/2,\mu-1/2)}(\cos\theta)
	\{\cos(\theta/2)\}^{\mu}
	\{\sin(\theta/2)\}^{\nu}
\end{equation}
for the Hamiltonian \eqref{eq:PTham2}. A similar relation like \eqref{eq:reproducing} holds again. Hence we obtain
\begin{equation}
\begin{aligned}
\label{eq:kernel_PT}
	\langle\theta\vert e^{-\beta H/\hbar}\vert\theta{'}\rangle
	=&\lim\limits_{N\to\infty}
	\sum_{n=0}^{\infty}\left\{\sqrt{\frac{\ 8\pi\ }{\lambda}\,}
	I_{\nu+n}\left(\frac{\ 4\ }{\ \lambda\ }\right)
	\exp\left(-\frac{\ 4\ }{\ \lambda\ }-\frac{\ \lambda\ }{\ 32\ }\right)
	\right\}^{N}
	\phi_{n}^{(\mu,\nu)}(\theta)\phi_{n}^{(\mu,\nu)}(\theta{'})\\
	=&
	\sum_{n=0}^{\infty}
	e^{-\beta E_{n}^{(\mu,\nu)}/\hbar}
	\phi_{n}^{(\mu,\nu)}(\theta)\phi_{n}^{(\mu,\nu)}(\theta{'})
\end{aligned}
\end{equation}
solely by means of path integral method.

We have thus succeeded in formulation and evaluation of path integrals for generalized P\"{o}schl-Teller potential as well as the symmetric one. We should also emphasize here that the relative phase factor $e^{\nu\pi i}$ in the partial kernels for the symmetric case tends to $-1$ if we set $\nu=1$, that is, our result reproduces the well-known minus sign in the odd times reflected component of the Feynman kernel for the free particle in a box because for $\nu=1$ the potential disappears. Interestingly, if we set $\nu$ to be an even integer, we observe that the phase factor becomes unity so that paths reflected odd times also contribute the kernel in an additive way. In this regard, the factor $-1$ for the amplitude of the free particle in a box is usually understood from the viewpoint of boundary condition at the endpoints. Our derivation of this factor is based on the consideration of the behavior of wave functions around boundaries not the requirement for kenrel to vanish there. However, we are convinced, from the eigenfunction expansion, that our prescription yields correct kernel that fulfills the boundary conditions.

\subsection{Systems on the half-line}
We proceed now to consider path integrals for systems on the half-line. Typical example is given by the radial Schr\"{o}dinger equation. If we define the inner product of wave functions by
$\int_{0}^{\infty}\psi^{*}(x)\phi(x)\,dx$, the Hamiltonian for such a system may have the form given by \eqref{eq:hamiltonian01} in which the potential will be written as
\begin{equation}
	V(x)=\frac{\ \hbar^{2}\nu(\nu-1)\ }{\ 2mx^{2}\ }+U(x)\quad(0<x<\infty),
\end{equation}
where $\nu=l+(D-1)/2$ being specified by the angular momentum $l(l+D-2)$ for the radial Hamiltonian in $D$-dimensional space.

It is natural to choose $f(x)$ to be given by $f(x)=x^{2}/(2a)$ which yields
\begin{equation}
	Q(x)=a\log\frac{\ x\ }{\ a\ },
\end{equation}
where $a$ is positive constant carrying the dimension of length. For the radial oscillator, whose potential being specified by $U(x)=m\omega^{2}x^{2}/2$, $a=\sqrt{\hbar/m\omega\,}$ will be a natural choice and $a=a_{\mathrm{B}}$, $a_{\mathrm{B}}=\hbar^{2}/(km)$ for the radial Coulomb system, specified by $U(x)=-k/x$($k>0$), will be suitable. The corresponding momentum operator $P$ reads
\begin{equation}
	P=-\frac{\ i\hbar\ }{\ 2a\ }\left(
	x\frac{\ d\ \ }{\ dx\ }+
	\frac{\ d\ \ }{\ dx\ }x\right)
\end{equation}
whose eigenfunction being given by
\begin{equation}
	\psi_{P}(x)=\frac{1}{\ \sqrt{2\pi\hbar (x/a)\,}\ }
	\exp\left(\frac{\ i\ }{\ \hbar\ }aP\log\frac{\ x\ }{\ a\ }\right).
\end{equation}
The completeness of the eigenvectors can be seen by calculating
\begin{equation}
	\int_{-\infty}^{\infty}\!\!\psi_{P}(x)\psi^{*}_{P}(x{'})\,dP=
	\frac{1}{\ \sqrt{xx{'}\,}\ }\delta(\log x-\log x{'})=
	\delta(x-x{'})
\end{equation}
for positive $x$ and $x{'}$.

Since we are aiming here to show the usefulness of \eqref{eq:kernel01},
let us restrict ourselves to the case of radial oscillator so that we can find its exact solution. Introducing a dimensionless variable $u$ by $u=x/a$, $a=\sqrt{\hbar/m\omega\,}$, we find the short time kernel \eqref{eq:kernel01} for this system should be given by
\begin{equation}
\label{eq:kernel08}
	{\mathcal K}(u,u{'};\epsilon)=
	\frac{1}{\ \sqrt{2\pi\lambda\ }\,}
	\exp\left[-\frac{1}{\ 2\lambda\ }uu{'}
	\left(\log\frac{\ u\ }{\ u{'}\ }\right)^{2}-
	\lambda U_{\mathrm{eff}}(u,u{'})
	\right],\quad
	\lambda\equiv\frac{\ \hbar\epsilon\ }{\ ma^{2}\ }=\omega\epsilon,
\end{equation}
in which the effective potential being given by
\begin{equation}
	U_{\mathrm{eff}}(u,u{'})=\frac{1}{\ 8uu{'}\ }+
	\frac{\ \nu(\nu-1)\ }{\ 2uu{'}\ }+
	\frac{\ 1\ }{\ 4\ }(u^{2}+u{'}^{2}).
\end{equation}
The first term of $U_{\mathrm{eff}}(u,u{'})$ has appeared through rewriting the Hamiltonian in terms of $P$. Again, the kernel above is normalized to fit integration with respect to $u$ instead of $x$. We may replace the last term in the effective potential by $uu{'}/2$ as the result of the geometric mean. It does not, however, affect the argument below. We therefore use $(u^{2}+u{'}^{2})/4$ as the harmonic potential in the short time kernel for convenience. See appendix on the detail of this ambiguity in the form of the potential term.

We first examine the contribution from the saddle point at $u-u{'}=0$ to obtain
\begin{equation}
\label{eq:k_p01}
	{\mathcal K}^{(\mathrm{e})}(u,u{'};\epsilon)=
	\frac{1}{\ \sqrt{2\pi\lambda\,}\ }\exp\left[
	-\frac{1}{\ 2\lambda\ }(u-u{'})^{2}-
	\frac{\ \lambda\ }{\ 2\ }\frac{\ \nu(\nu-1)\ }{uu{'}}-
	\frac{\ \lambda\ }{\ 4\ }(u^{2}+u{'}{}^{2})
	\right]
\end{equation}
whose exponent can be arranged as
\begin{equation}
	\frac{1}{\ 2\lambda\ }(u-u{'})^{2}+
	\frac{\ \lambda\ }{\ 2\ }\frac{\ \nu(\nu-1)\ }{uu{'}}+
	\frac{\ \lambda\ }{\ 4\ }(u^{2}+u{'}{}^{2})=
	\frac{1}{\ 2\lambda\ }\left(1+\frac{\ \lambda^{2}\ }{2}\right)
	(u^{2}+u{'}^{2})-\frac{\ uu{'}\ }{\lambda}+
	\frac{\ \lambda\ }{\ 2\ }\frac{\ \nu(\nu-1)\ }{uu{'}}.
\end{equation}
We can always modify the exponent of a time sliced path integral by adding higher order terms of $\lambda$. Therefore we can rewrite \eqref{eq:k_p01} as
\begin{equation}
\label{eq:k_p02}
	{\mathcal K}^{(\mathrm{e})}(u,u{'};\epsilon)=
	\frac{1}{\ \sqrt{2\pi\sinh\lambda\,}\ }\exp\left[
	-\frac{\cosh\lambda}{\ 2\sinh\lambda\ }(u^{2}+u{'}^{2})+
	\frac{uu{'}}{\ \sinh\lambda\ }-
	\frac{\ \sinh\lambda\ }{\ 2\ }\frac{\ \nu(\nu-1)\ }{uu{'}}
	\right]
\end{equation}
without changing the result of the time sliced path integral.
Since ${\mathcal K}^{(\mathrm{e})}(u,u{'};\epsilon)$ can now be identified with the asymptotic form of
\begin{equation}
\label{eq:k_p03}
	\frac{\sqrt{uu{'}\,}}{\ \sinh\lambda\ }\exp\left[
	-\frac{\ \coth\lambda\ }{\ 2\ }
	(u^{2}+u{'}{}^{2})\right]
	I_{\nu-1/2}\left(\frac{\ uu{'}\ }{\sinh\lambda}\right)
\end{equation}
for $uu{'}/\sinh\lambda\to\infty$, we can utilize this expression to consider the contribution from the reflected paths by the analytic continuation of the modified Bessel function to find
\begin{equation}
\label{eq:k_o01}
	{\mathcal K}^{(\mathrm{o})}(u,u{'};\epsilon)=e^{\nu\pi i}
	\frac{1}{\ \sqrt{2\pi\sinh\lambda\,}\ }\exp\left[
	-\frac{\cosh\lambda}{\ 2\sinh\lambda\ }(u^{2}+u{'}^{2})-
	\frac{uu{'}}{\ \sinh\lambda\ }+
	\frac{\ \sinh\lambda\ }{\ 2\ }\frac{\ \nu(\nu-1)\ }{uu{'}}
	\right].
\end{equation}
We can rewrite it as
\begin{equation}
\label{eq:k_o02}
	{\mathcal K}^{(\mathrm{o})}(u,u{'};\epsilon)=e^{\nu\pi i}
	\frac{1}{\ \sqrt{2\pi\lambda\,}\ }\exp\left[
	-\frac{1}{\ 2\lambda\ }(u+u{'})^{2}+
	\frac{\ \lambda\ }{\ 2\ }\frac{\ \nu(\nu-1)\ }{uu{'}}-
	\frac{\ \lambda\ }{\ 4\ }(u^{2}+u{'}{}^{2})
	\right]
\end{equation}
to make it clear that ${\mathcal K}^{(\mathrm{o})}(u,u{'};\epsilon)$
can be viewed as the contribution from the saddle point at $u+u{'}=0$.

It will be now evident that ${\mathcal K}^{(\mathrm{e})}(u,u{'};\epsilon)$ and
${\mathcal K}^{(\mathrm{o})}(u,u{'};\epsilon)$ are corresponding to the terms in the asymptotic from of the modified Bessel function
\begin{equation}
\begin{aligned}
	&I_{\nu-1/2}\left(\frac{\ uu{'}}{\ \sinh\lambda\ }\right)
	\sim
	\sqrt{\frac{\ \sinh\lambda\ }{\ 2\pi uu{'}\ }\,}\\
	&\times
	\left[
	\exp\left\{\frac{\ uu{'}}{\ \sinh\lambda\ }-
	\frac{\ \sinh\lambda\ }{\ 2\ }
	\frac{\ \nu(\nu-1)\ }{\ uu{'}\ }\right\}+
	e^{\nu\pi i}
	\exp\left\{-\frac{\ uu{'}}{\ \sinh\lambda\ }+
	\frac{\ \sinh\lambda\ }{\ 2\ }
	\frac{\ \nu(\nu-1)\ }{\ uu{'}\ }\right\}
	\right].
\end{aligned}
\end{equation}
Therefore the decomposition of the kernel ${\mathcal K}(u,u{'};\epsilon)$ into the sum of ${\mathcal K}^{(\mathrm{e})}(u,u{'};\epsilon)$ and ${\mathcal K}^{(\mathrm{o})}(u,u{'};\epsilon)$ explains the extension to the covering space from the original domain. We thus obtain
\begin{equation}
\label{eq:kernel08a}
	{\mathcal K}(u,u{'};\epsilon)=
	\frac{\sqrt{uu{'}\,}}{\ \sinh\lambda\ }\exp\left[
	-\frac{\ \coth\lambda\ }{\ 2\ }
	(u^{2}+u{'}{}^{2})\right]
	I_{\nu-1/2}\left(\frac{\ uu{'}\ }{\sinh\lambda}\right)
\end{equation}
as the sum of ${\mathcal K}^{(\mathrm{e})}(u,u{'};\epsilon)$ and ${\mathcal K}^{(\mathrm{o})}(u,u{'};\epsilon)$.
If we make use of the formula
\begin{equation}
	\int_{0}^{\infty}\!\!e^{-ax^{2}}xI_{\nu}(px)I_{\nu}(qx)\,dx=
\frac{1}{\ 2a\ }e^{(p^{2}+q^{2})/4a}I_{\nu}\left(\frac{pq}{\ 2a\ }\right)
\end{equation}
which is valid for $a>0$, we can verify that there holds
\begin{equation}
	\int_{0}^{\infty}\!\!
	{\mathcal K}(u,u{'},\epsilon)
	{\mathcal K}(u{'},u{''},\epsilon)\,du{'}=
	{\mathcal K}(u,u{''},2\epsilon).
\end{equation}
This proves that the form of the short time kernel given by \eqref{eq:kernel08a} is already exact and hence $\epsilon$ can be finite.
We then obtain ${\mathcal K}(u,u{'};\beta)$ in the same form as the one in \eqref{eq:kernel08a} by substituting $\omega\beta$ to $\lambda$ for finite $\beta$.

We may make use of a formula(see e.g. Ch. 4.17 in ref.\onlinecite{Lebedev})
\begin{equation}
	\sum_{n=0}^{\infty}\!\!\!\frac{\ n!L_{n}^{\alpha}(x)L_{n}^{\alpha}(y)\ }
	{\ \varGamma(n+\alpha+1)\ }t^{n}=
	\frac{e^{-(x+y)t/(1-t)}}{1-t}(xyt)^{-\alpha/2}
	I_{\alpha}\left(\frac{\ 2(xyt)^{1/2}\ }{1-t}\right)
\end{equation}
to find that the Euclidean kernel ${\mathcal K}(u,u{'};\beta)$ has an expansion
\begin{equation}
\label{eq:kernel09}
	{\mathcal K}(u,u{'};\beta)=
	\sum_{n=0}^{\infty}e^{-(2n+\nu+1/2)\lambda}
	\phi_{n}^{(\nu)}(u)\phi_{n}^{(\nu)}(u{'})
\end{equation}
in terms of the eigenfunction of the Hamiltonian. The explicit form of the eigenfunction $\phi_{n}^{(\nu)}(u)$ for this case is given by
\begin{equation}
	\phi_{n}^{(\nu)}(u)\equiv\sqrt{\frac{\ 2n!\ }
	{\ \varGamma(\nu+n+1/2)\ }\,}e^{-u^{2}/2}u^{\nu}
	L_{n}^{\nu-1/2}(u^{2})
\end{equation}
in terms of the Laguerre polynomial $L_{n}^{\nu-1/2}(u^{2})$.
The eigenvalue of the Hamiltonian is also obtained immediately from \eqref{eq:kernel09} to be $E_{n}^{(\nu)}=(2n+\nu+1/2)\hbar\omega$.
In this way, we have successfully formulated and solved path integral for the radial oscillator in an exact manner.

\section{Conclusion}

We have developed a method to obtain Lagrangian path integrals for systems defined on a finite interval as well as for systems on the half-line. The first step of our method is finding a suitable function of the position variable $x$ so that we are able to define a point transformation from $x$ to $Q(x)$ which covers all real values. This point transformation allows us to introduce the canonical momentum $P$ associated with $Q(x)$. It is the completeness of the eigenvectors of this new momentum operator that plays the role in our method to convert a Hamiltonian path integral into the corresponding Lagrangian one via the Gaussian integration. For systems on a finite interval or on the half-line, the kinetic term of the Lagrangian path integral thus formulated possesses multiple saddle points to give contributions to the path integral. We have taken them into account by extending the original domain to the covering space in obtaining suitable expressions for the short time kernel. In solving the path integral of a radial oscillator, we have succeeded formulating and carrying out the path integral entirely within the framework of path integral method. In this regard, we have to content ourselves to rely on some identities to convert short time kernels into the corresponding eigenfunction expansions for systems on a finite interval. It will be beautiful and useful if we could obtain short time kernels for these systems in a closed form by which we can validate the reproducing property of kernels in an exact manner. The phase factor $e^{\nu\pi i}$ appeared in the partial kernel for the symmetric P\"{o}schl-Teller potential generalizes the factor $-1$ for the odd times reflected amplitude of the free particle in a box. Our result clearly shows that the phase factor the Feynman kernel acquires upon reflections at boundaries depends on the dynamics of the system and therefore cannot be determined just by the consideration on the geometry of the system.

On the point transformation we have employed to formulate path integrals for systems with non-trivial geometry, we may fully utilize the completeness of eigenvectors of $Q=Q(x)$ as well as $P$ leaving from the viewpoint of path integrals for Feynman kernels in terms of the original variable $x$. This new pair of canonical variables is quite suitable for making consideration on the DK transformed path integral in the time sliced representation. Relations among solvable models, such as DK equivalence of the radial Coulomb path integral to that of the radial oscillator\cite{Sakoda2017}, can be deduced in a rigorous manner by formulating time sliced path integrals in terms of these new canonical variables. 

\section*{Acknowledgement}
The author would like to express his gratitude to Dr. Taro Kashiwa, a professor emeritus at Ehime university, for drawing his attention to the main subject of this paper.

\appendix
\section{Ambiguity in the form of the potential term of a path integral}

In this appendix, we consider the possibility of the different choice of the from of potential term in the short time kernel \eqref{eq:kernel08}.
If we evaluate the matrix element $\langle x\vert x^{2}\vert x{'}\rangle$ as
$xx{'}\langle x\vert x{'}\rangle$, the exponent of the kernel becomes
\begin{equation}
	-\frac{1}{\ 2\lambda\ }(u-u{'})^{2}-
	\frac{\ \lambda\ }{\ 2\ }\frac{\ \nu(\nu-1)\ }{uu{'}}-
	\frac{\ \lambda\ }{\ 2\ }uu{'}
\end{equation}
which can be arranged to yield
\begin{equation}
	-\frac{1}{\ 2\lambda\ }(u^{2}+u{'}^{2})-
	\frac{\ \lambda\ }{\ 2\ }\frac{\ \nu(\nu-1)\ }{uu{'}}+
	\left(1-\frac{\ \lambda^{2}\ }{2}\right)
	\frac{\ uu{'}\ }{\ \lambda\ }.
\end{equation}
We may set $v=(1-\lambda^{2}/2)^{1/2}u$ and write $u$ for $v$ again
to find that the expression above is equivalent to
\begin{equation}
\label{eq:exp02}
	-\frac{1}{\ 2\lambda\ }\left(1+\frac{\ \lambda^{2}\ }{2}\right)
	(u^{2}+u{'}^{2})-
	\frac{\ \lambda\ }{\ 2\ }\frac{\ \nu(\nu-1)\ }{uu{'}}+
	\frac{\ uu{'}\ }{\ \lambda\ }
\end{equation}
by discarding the irrelevant terms. Since the Jacobian of the change of variables from $u$ to $v$ converges to unity in the continuum limit, we can regard it as unity from very beginning. It is, therefore, clear that the different choice for the matrix element $\langle x\vert x^{2}\vert x{'}\rangle$
to be $xx{'}\langle x\vert x{'}\rangle$ does not affect the final form of the Euclidean kernel. A more generalized scheme for this matrix element will be given by $x^{1+\alpha}x{'}^{1-\alpha}\langle x\vert x{'}\rangle$. This is, however, expressed as
\begin{equation}
	x^{1+\alpha}x{'}^{1-\alpha}=
	xx{'}\left(\frac{\ x\ }{\ x{'}\ }\right)^{\alpha}
\end{equation}
to exhibit the fact that $\alpha$-dependent factor generates irrelevant terms in addition to unity in the exponent of a path integral. Therefore, it is equivalent to the symmetric one considered above.


\clearpage

\begin{thebibliography}{99}
\bibitem{Schrodinger46A9}E. Schro\"{o}dinger, Proc. Roy. Irish Acad. 46 A, 9 (1940).
\bibitem{Schrodinger46A183}E. Schro\"{o}dinger,Proc. Roy. Irish Acad. 46 A, 183 (1941).
\bibitem{Schrodinger47A}E. Schro\"{o}dinger,Proc. Roy. Irish Acad. 47 A, 53 (1941).
\bibitem{Dirac}P.M. Dirac, {\em The principles of Quantum Mechanics}, Oxford, (N.Y., 1935).
\bibitem{Infeld_Hull}I. Infeld and T.E. Hull, Rev. Mod. Phys. {\bf 23}, 21 (1951).
\bibitem{Gendenshtein}L. E. Gendenshtein, JETP Lett. {\bf 38}, 356 (1983).
\bibitem{Witten1981}E. Witten, Nucl. Phys. {\bf B188}, 513 (1981).
\bibitem{DKS}J. W. Dabrowska, A. Khare and U. P. Sukhatome, J. Phys. {\bf A 21}, L195 (1988).
\bibitem{CKS}F. Cooper, A. Khare and U. Sukhatme, Phys. Rep. {\bf 251}, 267 (1995).
\bibitem{CKSbook}F. Cooper, A. Khare and U. Sukhatme, {\em Supersymmetry In Quantum Mechanics}, World Scientific (Singapore, 2001).
\bibitem{Crum}M. M. Crum, Quart. J. Math. Oxford Ser. (2) {\bf 6}, 121 (1955), \texttt{arXiv:physics/9908019}.
\bibitem{Odake_Sasaki}S. Odake and R. Sasaki, J. Math. Phys. {\bf 47}, 102102 (2006).
\bibitem{DK79}I.~H. Duru and H. Kleinert, Phys. Lett. {\bf B84}, 185 (1979).
\bibitem{DK}I.~H. Duru and H. Kleinert, Fortschr. d. Phys. {\bf 30}, 401 (1982).
\bibitem{Inomata82} A. Inomata, Phys. Lett. {\bf A87}, 387 (1982).
\bibitem{HoInomata}R. Ho and A. Inomata, Phys. Rev. Lett. {\bf 48}, 231 (1982).
\bibitem{Inomata}A. Inomata, Phys. Lett. {\bf A101}, 253 (1984).
\bibitem{RingwoodDevreese}G.~A. Ringwood and J.~T. Devreese, J. Math. Phys. {\bf 21}, 1390 (1980).
\bibitem{BlanchardSirugue}P. Blanchard and M. Sirugue, J. Math. Phys. {\bf 22}, 1372 (1981).
\bibitem{Steiner}F. Steiner, Phys. Lett. {\bf A106}, 363 (1984).
\bibitem{PakSokmen}N.~K. Pak, and I. Soekmen, Phys. Rev. {\bf A30}, 1629 (1984).
\bibitem{Duru86}I.~H. Duru, Phys. Lett. {\bf A119}, 163 (1986).
\bibitem{ChetouaniChetouani}L. Chetouani and T.~F. Chetouani, J. Math. Phys. {\bf 27}, 2944 (1986).
\bibitem{YoungDeWitt-Morette}A. Young and C. DeWitt-Morette, Ann. Phys. (NY) {\bf 169}, 140 (1986).
\bibitem{Bohm_Junker}M. B\"{o}hm and G. Junker, J. Math. Phys. {\bf 29}, 1978 (1987).
\bibitem{Kleinert87} H. Kleinert, Phys. Lett. {\bf A120}, 361 (1987).
\bibitem{CastrigianoStaerk}D.~P.~L. Castrigiano and F. St\"{a}rk, J. Math. Phys. {\bf 30}, 2785 (1989).
\bibitem{KleinertBook}H. Kleinert, {\it Path integrals in quantum mechanics, statistics, and polymer physics\/}, 2nd edition, World Scientific (Singapore, 1995).
\bibitem{Kleinert98}H. Kleinert, Phys. Lett. {\bf A252}, 277 (1999).
\bibitem{Fujikawa}K. Fujikawa, Nucl. Phys. B {\bf 484}, 495 (1997).
\bibitem{Sakoda}S. Sakoda, Mod. Phys. Lett. {\bf A23}, 3057 (2008).
\bibitem{Sakoda2017}S. Sakoda, J. Math. Phys. {\bf 58}, 062111 (2017).

\bibitem{Rajeev1994}S.~G. Rajeev, S.~K. Rama and S. Sen, J. Math. Phys. {\bf 35}, 2259 (1994).
\bibitem{FKSF95_3232}K. Funahashi, T. Kashiwa, S. Sakoda and K. Fujii, J. Math. Phys. {\bf 36}, 3232 (1995).
\bibitem{FKSF95_4590}K. Funahashi, T. Kashiwa, S. Sakoda and K. Fujii, J. Math. Phys. {\bf 36}, 4590 (1995).
\bibitem{FKNS95}K. Funahashi, T. Kashiwa, S. Nima and S. Sakoda, Nucl. Phys. {\bf B453}, 508 (1995).
\bibitem{FKS96}K. Fujii, T. Kashiwa and S. Sakoda, J. Math. Phys. {\bf 37}, 567 (1996).

\bibitem{Schulman68}L.~S. Schulman, Phys. Rev. {\bf 176}, 1558 (1968).
\bibitem{Pauli1974}W. Pauli, {\em Pauli lectures on physics}, Dover edition, Vol. 5, ed. C.~P. Enz, Dover publications(New York, 2000).
\bibitem{Laidlaw_DeWitt}M.~G.~G. Laidlaw and C. DeWitt, Phys. Rev. {\bf D3}, 1375 (1971).
\bibitem{DeWitt_Morette}C. DeWitt-Morette, A. Maheshwari and B. Nelson, Phys. Rep. {\bf 50}, 255 (1979). 
\bibitem{Janke_Kleinert}W. Janke and H. Kleinert, Lett. al Nuovo Cimento {\bf 25}, 297 (1979).
\bibitem{Marinov80}M.~S. Marinov, Phys. Rep. {\bf 60}, 1 (1980).
\bibitem{Inomate_Singh}A. Inomata and V.~A. Singh, Phys. Lett. {\bf A80}, 105 (1980).
\bibitem{Fujikawa2008}K. Fujikawa, Prog. Theor. Phys. {\bf 120}, 181 (2008).
\bibitem{CMS1980}T.~E. Clark, R. Menikoff, and D.~H. Sharp, Phys. Rev. {\bf D22}, 3012 (1980).
\bibitem{Farhi_Gutmann}E. Farhi and S. Gutmann, Int. J. Mod. Phys. {\bf A05}, 3029 (1990).
\bibitem{Gervai_Jevicki}J.~L. Gervais and A. Jevicki, Nucl. Phys. {\bf B110}, 93 (1976).
\bibitem{Omote1977}M. Omote, Nucl. Phys. {\bf B120}, 325 (1977).
\bibitem{FukutakaKashiwa}H. Fukutaka and T. Kashiwa, Ann. Phys. (NY) {\bf 185}, 301 (1983).
\bibitem{Ohnuki_Watanabe2003}Y. Ohnuki and S. Watanabe, J. Anal. Appl. {\bf 1}, 193 (2003).
\bibitem{S.Ohya}S. Ohya, a comment given for a talk at Nihon University and in private communications. To reflect the symmetry of the Hamiltonian under the exchange $\nu$ by $1-\nu$, if we do not restrict $\nu$ to satisfy $\nu\ge1/2$, we should replace $\nu$ by $\vert{\nu-1/2}\vert+1/2$ here and in the following. The same will apply for the parameter $\mu$ in the next model and $\nu$ again in the radial harmonic oscillator.
\bibitem{Watson}G. N. Watson, {\em Theory of Bessel functions}, 2nd edition, Cambridge University Press(New York, 1966).
\bibitem{Sakoda2018b}S. Sakoda, in preparation.
\bibitem{Lebedev}N.~N. Lebedev, {\em SPECIAL FUNCTIONS AND THEIR APPLICATIONS}, Dover edition, tr. and ed. by R.~A. Silverman, Dover publications(New York, 1972).

\end{thebibliography}
\end{document}